\documentclass{pasj00}
\usepackage{rotating}
\usepackage{ccaption}

\begin{document}
\SetRunningHead{Doi et al.}{VLBI Detections of Radio-Loud BAL Quasars}
\Received{2008/12/13}%{yyyy/mm/dd}
\Accepted{2009/10/18}%{yyyy/mm/dd}

\title{VLBI Detections of Parsec-Scale Nonthermal Jets\\ in Radio-Loud Broad Absorption Line Quasars}

\author{
Akihiro \textsc{Doi},\altaffilmark{1,2,3}
Noriyuki \textsc{Kawaguchi},\altaffilmark{4,5}
Yusuke \textsc{Kono},\altaffilmark{4}
Tomoaki \textsc{Oyama},\altaffilmark{6}
Kenta \textsc{Fujisawa},\altaffilmark{3}\\
Hiroshi \textsc{Takaba},\altaffilmark{7}
Hiroshi \textsc{Sudou},\altaffilmark{7}
Ken-ichi \textsc{Wakamatsu},\altaffilmark{7}
Aya \textsc{Yamauchi},\altaffilmark{6,8}\\
Yasuhiro \textsc{Murata},\altaffilmark{1,2}
Nanako \textsc{Mochizuki},\altaffilmark{1}
Kiyoaki \textsc{Wajima},\altaffilmark{3}
Toshihiro \textsc{Omodaka},\altaffilmark{9}\\
Takumi \textsc{Nagayama},\altaffilmark{10}
Naomasa \textsc{Nakai},\altaffilmark{8}
Kazuo \textsc{Sorai},\altaffilmark{11}
Eiji \textsc{Kawai},\altaffilmark{12}
Mamoru \textsc{Sekido},\altaffilmark{12}\\
Yasuhiro \textsc{Koyama},\altaffilmark{12}
the VLBI group at Geographical Survey Institute,\altaffilmark{13}\\
Shoichiro \textsc{Asano},\altaffilmark{14} and 
Hisao \textsc{Uose}\altaffilmark{15}
}

\altaffiltext{1}{The Institute of Space and Astronautical Science, Japan Aerospace Exploration Agency,\\ 3-1-1 Yoshinodai, Sagamihara, Kanagawa 229-8510}
\altaffiltext{2}{Department of Space and Astronautical Science, The Graduate University for Advanced Studies,\\ 3-1-1 Yoshinodai, Sagamihara, Kanagawa 229-8510}
\altaffiltext{3}{Faculty of Science, Yamaguchi University, 1677-1 Yoshida, Yamaguchi, Yamaguchi 753-8512}
\altaffiltext{4}{National Astronomical Observatory of Japan, 2-21-1 Osawa, Mitaka, Tokyo 181-8588}
\altaffiltext{5}{Department of Astronomical Science, Graduate University for Advanced Studies,\\ 2-21-1 Osawa, Mitaka, Tokyo 181-8588}
\altaffiltext{6}{Mizusawa VERA Observatory, 2-12 Hoshigaoka, Mizusawa, Oshu, Iwate 023-0861}
\altaffiltext{7}{Faculty of Engineering, Gifu University, 1-1 Yanagido, Gifu 501-1193}
\altaffiltext{8}{Institute of Physics, University of Tsukuba, 1-1-1 Ten-noudai, Tsukuba, Ibaraki 305-8571}
\altaffiltext{9}{Faculty of Science, Kagoshima University, 1-21-30 Korimoto, Kagoshima, Kagoshima 890-0065}
\altaffiltext{10}{Graduate School of Science and Engineering, Kagoshima University,\\ 1-21-35 Korimoto, Kagoshima, Kagoshima 890-0065}
\altaffiltext{11}{Division of Physics, Graduate School of Science, Hokkaido University, N10W8, Sapporo, Hokkaido 060-0810}
\altaffiltext{12}{Kashima Space Reserch Center, National Institute of Information and Communications Technology,\\ 893-1 Hirai, Kashima, Ibaraki 314-8501}
\altaffiltext{13}{Geographical Survey Institute, 1 Kitasato, Tsukuba, Ibaraki, 305-0811}
\altaffiltext{14}{National Institute of Infomatics, 2-1-2, Hitotsubashi, Chiyoda-ku, Tokyo 101-8430}
\altaffiltext{15}{NTT Advanced Technology Corporation, 3-9-11 Midori-cho, Musashino-shi, Tokyo, 180-8585}


\maketitle

\begin{abstract}
We conducted radio detection observations at 8.4~GHz for 22~radio-loud broad absorption line~(BAL) quasars, selected from the Sloan Digital Sky Survey~(SDSS) Third Data Release, by a very-long-baseline interferometry~(VLBI) technique.  The VLBI instrument we used was developed by the Optically ConnecTed Array for VLBI Exploration project~(OCTAVE), which is operated as a subarray of the Japanese VLBI Network~(JVN).  We aimed at selecting BAL quasars with nonthermal jets suitable for measuring their orientation angles and ages by subsequent detailed VLBI imaging studies to evaluate two controversial issues of whether BAL quasars are viewed nearly edge-on, and of whether BAL quasars are in a short-lived evolutionary phase of quasar population.  We detected 20 out of 22 sources using the OCTAVE baselines, implying brightness temperatures greater than 10$^{5}$~K, which presumably come from nonthermal jets.  Hence, BAL outflows and nonthermal jets can be generated simultaneously in these central engines.  We also found four inverted-spectrum sources, which are interpreted as Doppler-beamed, pole-on-viewed relativistic jet sources or young radio sources: single edge-on geometry cannot describe all BAL quasars.  We discuss the implications of the OCTAVE observations for investigations for the orientation and evolutionary stage of BAL quasars.  
\end{abstract}

\section{Introduction}
Broad absorption line (BAL) quasars are a subclass of active galactic nuclei~(AGNs) with rest-frame ultra-violet spectra showing absorption troughs displaced blueward from the corresponding emission lines in the high-ionization transitions of C$_\mathrm{IV}$, Si$_\mathrm{IV}$, N$_\mathrm{V}$, and O$_\mathrm{IV}$, and occasionally in low-ionization transitions, e.g., Mg$_\mathrm{II}$ and Al$_\mathrm{III}$ \citep{Weymann_etal.1991}.  The absorption troughs are broader than 2000~km~s$^{-1}$, sometimes as broad as $\sim0.1c$, and are presumably due to the intervening components of the outflow originating in the activity of central engines.  The fact that the most luminous quasars showing BALs more frequently \citep{Ganguly_etal.2007} is consistent with the strong radiation-pressure-driven outflows suggested by simulation-based studies for accretion phenomena (e.g., \cite{Proga_etal.2000,Ohsuga2007}).  The observed maximum velocity of absorption as function of luminosity has an upper envelope, which can be interpreted as the terminal velocity of radiative driven wind \citep{Ganguly_etal.2007,Laor&Brandt2002}.  

The intrinsic percentage of quasars with BALs is $\sim$20\% (e.g., \cite{Hewett&Foltz2003}).  This percentage means that the BAL phenomenon takes one of major roles in quasars'~activities.  However, the principal parameter determining the finding of BAL features is still unknown.  In the most widely accepted scenario, this percentage represents the covering factor of an outflowing BAL wind, which is preferentially equatorial, and the BAL features can be observed when the accretion disk is almost edge-on to the line of sight, based on spectropolarimetric measurements \citep{Goodrich&Miller1995,Cohen_etal.1995} and a theoretical disk wind model \citep{Murray_etal.1995}.  However, some radio observations provided a counterargument to this paradigm.  \citet{Zhou_etal.2006} and \citet{Ghosh&Punsly2007} found several BAL quasars with rapid radio variability that indicated very high brightness temperatures, which require Doppler beaming on jets with inclinations of less than \timeform{35D}, i.e., a nearly face-on view of the accretion disk.  \citet{Becker_etal.2000} found that about one-third of the radio-detected BAL quasars showed flat radio spectra ($\alpha>-0.5$, $S_\nu \propto \nu^{\alpha}$), preferring pole-on jets because this geometry tends to make radio sources core-dominated by significant Doppler effect only on nuclear jets.  Also in optical spectropolarimetry, \citet{Brotherton_etal.2006} found an electric vector nearly parallel to a large-scale jet axis, implying that BAL outflow is not equatorial, in a Fanaroff-Riley Class II~(FR~II) radio galaxy as a BAL quasar.  Thus, a pole-on outflow would be necessary for at least some of the known radio-emitting BAL quasars.  These results support an alternative proposal that BALs are not closely related to inclinations, and may be associated with a relatively short-lived (possibly episodic) evolutionary phase with a large BAL wind-covering fraction (e.g., \cite{Briggs_etal.1984,Gregg_etal.2000}).  \citet{Gregg_etal.2006} pointed out the rarity of FR~II/BAL quasars and their observed anticorrelation between the balnicity index and radio loudness, and suggested that these properties can be explained naturally by an evolutionary scheme.  \citet{Montenegro-Montes_etal.2008} indicated that many radio-emitting BAL quasars share several radio properties common to young radio sources like Compact Steep Spectrum~(CSS) or Gigahertz-Peaked Spectrum~(GPS) sources.  Thus, `inclination angle' and `evolutionary phase' are two of the most important aspects in understanding BAL quasars.  

Very-long-baseline interferometry~(VLBI) instruments in radio wavelengths provide exclusive and crucial opportunities to obtain spatial information about AGNs at milli-arcsecond~(mas) scales by direct measurement, which should also be useful for investigating BAL quasars.  The inclination can be resolved by determining the viewing angle of the jet axis, which is supposed to be perpendicular to the accretion disk.  Using the framework of Doppler beaming effects, jet axes for many AGNs have been estimated by measurements of jet asymmetry~(advancing speed, brightness, jet length, etc.) by VLBI imaging.  Age as a radio source can be also estimated by measurements of its apparent linear size and expanding speed, or the age of relativistic electrons appearing on synchrotron spectrum (e.g., \cite{Nagai_etal.2006} and references therein).  VLBI observations for BAL quasars have recently begun to try to study such phenomena.  \citet{Jiang&Wang2003} observed three BAL quasars with the European VLBI Network~(EVN) at 1.6~GHz and suggested that the jet of J1556+3517 was possibly viewed from nearly pole-on because of a flat spectrum and unresolved core, while that of J1312+2319 may be far from pole-on because of the two-sided structure, and the inclination of J0957+2356 was unclear because of an unresolved steep-spectrum compact source.  \citet{Liu_etal.2008} observed for eight BAL quasar sample, including both of LoBALs and HiBALs and both of steep- and flat-spectrum sources, with the EVN$+$Multi-Element Radio Linked Interferometer Network~(MERLIN) at 1.6~GHz.  High brightness temperatures and linear polarization in their core components implied a synchrotron origin for the radio emission.  No systematic difference was found in the radio morphology or polarization properties between their limited number of LoBAL/HiBAL or steep/flat-spectrum sources.  \citet{Kunert-Bajraszewska&Marecki2007} observed 1045+352 with the US Very Long Baseline Array~(VLBA) at 1.7, 5, and 8.4~GHz, and found complicated radio morphology and a projected linear size of only 2.1~kpc, which suggests it might be in an early stage, and consistent with the evolutionary scenario.  These investigations were still inconclusive because the detected jet structures, spatial resolutions, and frequency coverages were inadequate to definitively determine jet properties, and also because of still a small number of objects to conclude the natures of BAL quasars.

In this paper, we report our VLBI observations of 22~BAL quasars at 8.4~GHz by direct measurement at the mas resolution, corresponding to the parsec scale at the distance to these sources.  Our aim is to find BAL quasars that have jets suitable for determining their orientation angles and ages from jet properties for future detailed VLBI imaging studies.  In Section~\ref{section:sample}, we describe our selection processes, and we present our observations and data reduction procedures in Section~\ref{section:observationanddatareduction}.  We present the observational results in Section~\ref{section:result}, and discuss their implications in Section~\ref{section:discussion}.  In Section~\ref{section:summary}, we summarize the outcome of our investigation.  Throughout this paper, a flat cosmology is assumed, with $H_0=70$~km~s$^{-1}$~Mpc$^{-1}$, $\Omega_\mathrm{M}=0.3$, and $\Omega_\mathrm{\Lambda}=0.7$ \citep{Spergel_etal.2003}.

\begin{table*}
\centering
\rotatebox[origin=c]{90}{
\begin{minipage}[c]{1.25\textwidth}
\caption{Optically-selected, radio-flux-limited BAL quasar sample for OCTAVE observation.}\label{table1}
\begin{center}
\begin{tabular}{llclccrccc} \hline\hline
\multicolumn{1}{c}{SDSS name} & \multicolumn{1}{c}{FIRST name} & $r_\mathrm{FIRST}^\mathrm{SDSS}$ & \multicolumn{1}{c}{$z$} & $l$ & BAL type & \multicolumn{1}{c}{$M_i^\mathrm{SDSS}$} & $I^\mathrm{FIRST}_\mathrm{1.4GHz}$ & $S^\mathrm{FIRST}_\mathrm{1.4GHz}$ & $\log{R*}$ \\
\multicolumn{1}{c}{} & \multicolumn{1}{c}{} & (arcsec) & \multicolumn{1}{c}{} & (pc mas$^{-1}$) &  & \multicolumn{1}{c}{(mag)} & (mJy beam$^{-1}$) & (mJy) &  \\
\multicolumn{1}{c}{(1)} & \multicolumn{1}{c}{(2)} & (3) & \multicolumn{1}{c}{(4)} & (5) & (6) & \multicolumn{1}{c}{(7)} & (8) & (9) & (10) \\\hline
SDSS J004323.43$-$001552.4 & FIRST J004323.8$-$001548 & 7.6  & 2.798  & 7.9  & H & $-$27.94 & 103  & 115  & 2.7  \\
SDSS J021728.62$-$005227.2 & FIRST J021728.6$-$005226 & 0.3  & 2.463  & 8.1  & H & $-$25.98 & 212  & 218  & 3.7  \\
SDSS J075628.24$+$371455.6 & FIRST J075628.2$+$371455 & 0.3  & 2.514  & 8.1  & HL? & $-$26.10 & 239  & 247  & 3.7  \\
SDSS J080016.09$+$402955.6 & FIRST J080016.0$+$402955 & 0.5  & 2.021  & 8.4  & Hi & $-$26.80 & 190  & 200  & 3.1  \\
SDSS J081534.16$+$330528.9 & FIRST J081534.1$+$330529 & 0.5  & 2.426  & 8.1  & nH & $-$27.04 & 328  & 342  & 3.4  \\
SDSS J092824.13$+$444604.7 & FIRST J092824.1$+$444604 & 0.0  & 1.904  & 8.4  & nHi & $-$27.16 & 156  & 162  & 2.8  \\
SDSS J100515.98$+$480533.2 & FIRST J100515.9$+$480533 & 0.2  & 2.372  & 8.2  & Hi & $-$28.29 & 206  & 209  & 2.7  \\
SDSS J101329.92$+$491840.9 & FIRST J101329.9$+$491841 & 0.2  & 2.201  & 8.3  & nHi & $-$26.86 & 267  & 269  & 3.3  \\
SDSS J101827.85$+$053030.0 & FIRST J101827.8$+$053029 & 0.2  & 1.938  & 8.4  & Hi & $-$26.62 & 284  & 297  & 3.3  \\
SDSS J102027.20$+$432056.2 & FIRST J102027.2$+$432056 & 0.2  & 1.962  & 8.4  & nHi & $-$26.74 & 108  & 110  & 2.9  \\
SDSS J103038.38$+$085324.9 & FIRST J103038.3$+$085324 & 0.6  & 1.750  & 8.5  & nHi & $-$26.57 & 108  & 172  & 3.0  \\
SDSS J104257.58$+$074850.5 & FIRST J104257.5$+$074850 & 0.3  & 2.665  & 8.0  & H & $-$27.32 & 374  & 382  & 3.5  \\
SDSS J105726.62$+$032448.0 & FIRST J105726.6$+$032448 & 0.5  & 2.832  & 7.3  & H & $-$26.90 & 138  & 157  & 3.2  \\
SDSS J110344.53$+$023209.9 & FIRST J110344.5$+$023209 & 0.2  & 2.514  & 8.1  & H & $-$27.63 & 163  & 166  & 2.9  \\
SDSS J111914.32$+$600457.2 & FIRST J111914.3$+$600457 & 0.1  & 2.646  & 8.0  & nH & $-$28.97 & 186  & 192  & 2.5  \\
SDSS J115944.82$+$011206.9 & FIRST J115944.8$+$011206 & 0.2  & 2.000  & 8.4  & Hi & $-$28.40 & 267  & 268  & 2.6  \\
SDSS J122343.16$+$503753.4 & FIRST J122343.1$+$503753 & 0.1  & 3.488  & 7.3  & H & $-$29.21 & 222  & 229  & 2.7  \\
SDSS J122836.92$-$030439.2 & FIRST J122836.9$-$030438 & 0.7  & 1.801  & 8.4  & nHi & $-$26.53 & 144  & 149  & 3.0  \\
SDSS J140507.80$+$405657.8 & FIRST J140507.7$+$405658 & 0.3  & 1.993  & 8.4  & Hi & $-$26.31 & 206  & 214  & 3.3  \\
SDSS J143243.29$+$410327.9 & FIRST J143243.3$+$410328 & 0.4  & 1.970  & 8.4  & nHi & $-$27.84 & 257  & 262  & 2.8  \\
SDSS J151005.88$+$595853.3 & FIRST J151005.4$+$595856 & 4.3  & 1.720  & 8.5  & nHi & $-$27.05 & 182  & 307  & 3.1  \\
SDSS J152821.65$+$531030.4 & FIRST J152821.6$+$531030 & 0.4  & 2.822  & 7.8  & H & $-$26.37 & 172  & 183  & 3.6  \\\hline
\end{tabular}
\end{center}
\begin{flushleft}
{\footnotesize Col.~(1) SDSS source name; Col.~(2) Radio counter part from FIRST; Col.~(3) Difference between SDSS--FIRST positions; Col.~(4) Redshift; Col.~(5) Linear scale corresponding to 1~mas; Col.~(6) BAL subtype, listed in \citet{Trump_etal.2006}.  The code ``n'' denotes a relatively narrow trough; ``HL'' denotes a HiBAL in which broad ($\geq$1000~km~s$^{-1}$) low-ionization absorption is also seen; ``Hi'' denotes a HiBAL-only object, in which broad low-ionization absorption is not seen even though Mg$_\mathrm{II}$ is within the spectral coverage; and ``H'' denotes a HiBAL in which the Mg$_\mathrm{II}$ region is not within the spectral coverage of SDSS or has a very low signal-to-noise ratio and so whether or not the object is a LoBAL as well as a HiBAL is unknown; Col.~(7) $i$-band absolute magnitude listed in \citet{Trump_etal.2006} and calculated using a flat cosmology of $H_0=70$~km~s$^{-1}$~Mpc$^{-1}$, $\Omega_\mathrm{M}=0.27$, and $\Omega_\mathrm{\Lambda}=0.73$ \citep{Spergel_etal.2003}; Col.~(8) 1.4~GHz peak intensity from FIRST data with a $\sim\timeform{5"}$ resolution; Col.~(9) 1.4~GHz flux density from FIRST data; Col.~(10) Radio loudness, the ratio of radio-to-optical flux densities, $R* = f_\mathrm{5GHz}/f_\mathrm{2500\AA}$ (e.g., \cite{Stocke_etal.1992}), which were calculated from $z$, $S^\mathrm{FIRST}_\mathrm{1.4GHz}$, $M_i$, and the assumption of a radio spectral index of $-0.5$ and an optical spectral index of $-0.5$.}
\end{flushleft}
\end{minipage}
} 
\end{table*}

\section{Sample}\label{section:sample}
The Sloan Digital Sky Survey~(SDSS) Third Data Release~(DR3; \cite{Abazajian_etal.2005}) cataloged 46,420~quasars \citep{Schneider_etal.2005}, including 4784~BAL quasars~\citep{Trump_etal.2006}.  In the catalog of the Very Large Array~(VLA) 1.4-GHz Faint Images of the Radio Sky at Twenty cm~(FIRST) survey \citep{Becker_etal.1995}, we searched for radio counterparts with $>$1~mJy located within \timeform{10''} of these SDSS BAL quasars.  This search radius was significantly large relative to the position uncertainties of the SDSS and FIRST ($\sim$\timeform{1''}), in order to rescue even moderately resolved lobe-dominated radio sources.  Of all SDSS BAL quasars, 91.4\% (4374/4784) were in the regions of the sky covered by the FIRST survey, and we found 492 radio sources corresponding to 11.2\% of SDSS BAL quasars.  Most of them were unresolved at a spatial resolution of $\sim$\timeform{5''} of the FIRST, suggesting intrinsic sizes of less than \timeform{1''} at the FIRST image sensitivity (see also \cite{Zhou_etal.2006}).  We selected out 23 sources with $>$100~mJy.  Except for three sources, these radio sources were found with separations of $<$\timeform{1''}.  FIRST~J004323.8$-$001548 and FIRST~J151005.4$+$595856 showed separation angles of \timeform{7.6''} and \timeform{4.3''} from their optical positions, respectively.  The two sources had significantly resolved radio structures, which provide the relatively large separations; FIRST~J004323.8$-$001548 had been already identified as an FR~II--BAL quasar by \cite{Gregg_etal.2006}).  FIRST~J103434.4$+$592445 was found at \timeform{9.9''} from SDSS~J103433.49$+$592452.3; this was probably a mis-identification because an another optical source, SDSS J103434.20+592445.8 (a candidate of gravitational lensing), was a much more likely source for the FIRST source.  Hence, we removed this source from the list.  Finally, our optically-selected, radio-flux-limited sample for the present VLBI observations consists of 22 sources~(Table~\ref{table1}).  

The redshift range is 1.72--3.488 with a median of $\sim2.2$ for the 22~selected BAL quasars.  The luminosity and angular distance are 17456~Mpc and 1705~Mpc at $z=2.2$, respectively.  An angular scale of 1~mas corresponds to a linear scale of 8.3~pc at $z=2.2$ (see Table~\ref{table1} for individual sources).  We determined radio loudness according to the definition of the ratio of radio-to-optical flux densities, $R* = f_\mathrm{5GHz}/f_\mathrm{2500\AA}$ (e.g., \cite{Stocke_etal.1992}), which was calculated from the redshift $z$, the FIRST flux density, and the SDSS $i$-band absolute magnitude, assuming $\alpha = -0.5$ for the radio and optical spectral indices (see Table~\ref{table1}).  All of our samples are very radio-loud ($\log{R*} \geqq 2.5$; cf.~the extensive study of radio-loudness for FIRST/SDSS sources by \cite{Ivezic_etal.2002}).  As this definition may somewhat exaggerate the radio loudness of BAL quasars with substantial optical reddening.  The largest extinction $A_i=0.22$~mag is estimated from $E(B-V)=0.114$ on SDSS J080016.09$+$402955.6 in the sample, and exaggerate the radio loudness by $\log{\Delta R*} \approx 0.1$.  It is also useful to consider the rest-frame monochromatic radio luminosity, where $L_\mathrm{5GHz}>10^{25}$~W~Hz$^{-1}$ distinguishes radio quasars (e.g., \cite{Miller_etal.1993}).  Our samples have $L_\mathrm{5GHz}\sim10^{27}$--10$^{28}$~W~Hz$^{-1}$ assuming $\alpha=-0.5$, are also diagnosed with radio-loud objects.

\section{Observations and data analysis}\label{section:observationanddatareduction}
We observed the selected 22~BAL quasars using an optical-fiber-linked real-time VLBI instrument constructed by the Optically ConnecTed Array for VLBI Exploration project~(OCTAVE; \cite{Kawaguchi2008}) and operated as a subarray of the Japanese VLBI Network~(JVN; \cite{Fujisawa2008}).  The OCTAVE connects six radio telescopes, the Usuda~64~m, Nobeyama~45~m, Kashima~34~m, Yamaguchi~32~m, Tsukuba~32~m, and Gifu~11~m, using optical fibers at 2.4~Gbps, and offers mJy-level fringe detection sensitivities.  Its high baseline sensitivities are crucial for efficient exploration of the VLBI-detectability of a large number of weak radio populations, such as BAL quasars, in preparation for detailed imaging studies.

\begin{table*}[hbt]
\centering
\begin{minipage}[c]{0.89\textwidth}
\caption{Observing sessions.}\label{table2}
\begin{center}
\begin{tabular}{llccccl} \cline{3-6}\hline\hline
\multicolumn{1}{c}{Date} & \multicolumn{1}{c}{Telescope} & \multicolumn{4}{c}{Target} & Flux reference \\
\multicolumn{1}{c}{(1)} & \multicolumn{1}{c}{(2)} & \multicolumn{4}{c}{(3)} & \multicolumn{1}{c}{(4)} \\\hline
2007 Nov 04 & UYT & 0756+371 & 0800+402 & 0815+330 & 0928+444 & DA 193 (4.39 Jy) \\
(14:00--16:30UT) &  & 1005+480 & 1013+491 & 1020+432 & 1119+600 &  \\
2007 Dec 02 & UYTK & 0043-001 & 0217-005 & 1223+503 & 1405+405 & NRAO 512 (1.62 Jy) \\
(05:40--12:00UT) &  & 1432+410 & 1510+595 & 1528+531 &  &  \\
2008 Feb 09 & UYT & 1018+053 & 1030+085 & 1042+074 & 1057+032 & OJ 287 (2.55 Jy) \\
(19:00--20:30UT) &  & 1103+023 & 1159+011 & 1228-030 &  &  \\\hline
\end{tabular}
\end{center}
\begin{flushleft}
{\footnotesize Col.~(1) Observation date (and time); Col.~(2) Telescopes used: ``U'' denotes Usuda 64~m, ``Y'' denotes Yamaguchi 32~m, ``T'' denotes Tsukuba 32~m, and ``K'' denotes Kashima 34~m; Col.~(3) Truncated name of target source; Col.~(4) Flux calibrator used for amplitude scaling.}
\end{flushleft}
\end{minipage}
\end{table*}

Observations were performed in three sessions~(Table~\ref{table2}) using the Usuda~64~m, Kashima~34~m, Yamaguchi~32~m, and Tsukuba~32~m telescopes, which had 8.4-GHz receiving systems in the OCTAVE array.  Radio signals of left-hand circular polarization were received and amplified with cooled low noise amplifiers~(LNAs) on each antenna.  An analog bandpass filter put through a bandwidth of 512~MHz (8192--8704~MHz), which was sampled into 2~bits (4~levels) at 1~Gsps using the ADS-1000 sampler.  The sampled data were transmitted directly through the optical fibers to dedicated real-time processing correlator units at the National Astronomical Observatory of Japan~(NAOJ), Mitaka, Tokyo.  The correlated data were converted into the Flexible Image Transport System~(FITS) format for analysis using the Astronomical Image Processing System~(AIPS; \cite{Greisen2003}) software.  We were able to obtain all final results within an hour after the end of an observation.

The scan period for each target was 10~minutes including an antenna slew, resulting in on-source integration of $\sim$8.0--9.5-minutes.  One short scan on a flux calibrator, which is an almost perfectly point radio source in the OCTAVE baselines, was also performed during each session (see Table~\ref{table2}).  The flux densities of the flux calibrators were also measured using a single-dish mode on the Yamaguchi 32~m telescope within several days before or after the VLBI observations.  The flux-calibrator scan determined antenna-gain ratios relative to each telescope and an absolute amplitude scaling factor at the time of the scan.  System-noise temperatures were monitored at each antenna to correct the time variation of sensitivities.  Such a method can achieve amplitude calibrations with a dispersion of $<$5\% for antenna-gain ratios and an uncertainty of $\sim$10\% for an absolute flux scaling, according to past Japanese VLBI Network observations at 8.4~GHz (e.g., \cite{Doi_etal.2006, Doi_etal.2007}).  The visibility data, which had been fringe-fitted and bandpass-calibrated, were averaged over frequency.  

We have performed non-imaging analysis for these calibrated data.  We simply applied a model-fitting to the visibility amplitudes of each baseline with a structure model of a Gaussian profile.  Free parameters of the Gaussian were the peak amplitude and full width at half maximum~(FWHM).  The OCTAVE had baselines in the range of 1.5--25~M$\lambda$, resulting in a sensitivity to sources' FWHM sizes ranging from $\sim$2 to 30--100~mas.  Because all of the baselines are almost in the east-west direction, our one-scan snapshot for each target had sensitivity to source's brightness profile roughly in one direction only.  The calibration uncertainty was less than 5\% in terms of amplitude gains relative to each antenna, and baseline sensitivities of 0.4--1.4~mJy.  To avoid amplitude biases due to low signal-to-noise ratios, we time-averaged visibilities with time intervals sufficient to improve the signal-to-noise ratios to more than 10 in the cases of weak sources.

The results of the model-fitting are listed in Table~\ref{table3}.  The source sizes of targets detected in more than one baseline could be measured straightforwardly.  For FIRST~J004323.8$-$001548 and FIRST~J080016.0$+$402955, we could not determine the source sizes because no fringe was detected in any baseline.  For FIRST~J021728.6$-$005226 and FIRST J152821.6$+$531030, only the shortest baselines could detect fringes, which provided lower limits of the source sizes, and also provided upper limits with the possibly reliable assumption that their total flux densities were equivalent to their FIRST peak intensities.  For FIRST~J100515.9$+$480533, model-fitting cannot be done because its visibility profile was quite different from a simple Gaussian.  Hence, we determined only an upper limit to its source size based on the detections with these baselines.  For FIRST J111914.3$+$600457, no model-fitting could be applied because data from only one baseline were available, due to trouble with the third antenna at that time.  We determined an upper limit to its source size in the same manner as the case of FIRST~J100515.9$+$480533.

\begin{table*}
\caption{Results of OCTAVE observations for 22 BAL quasars.}\label{table3}
\begin{center}
\begin{tabular}{lrrcccccc} \hline\hline
\multicolumn{1}{c}{Source} & \multicolumn{1}{c}{$S^\mathrm{cor}_\mathrm{8.4GHz}$} & \multicolumn{1}{c}{$B_{uv}$} & $S^\mathrm{fit}_\mathrm{8.4GHz}$ & $\phi^\mathrm{fit}_\mathrm{FWHM}$ & $\phi^\mathrm{fit}_\mathrm{FWHM}$ & $\log{(T_\mathrm{B})}$ & $I^\mathrm{FIRST}_\mathrm{1.4GHz}$ & $\alpha(I^\mathrm{FIRST}_\mathrm{1.4GHz} - S_\mathrm{8.4GHz}^\mathrm{cor-max})$ \\
\multicolumn{1}{c}{} & \multicolumn{1}{c}{(mJy)} & \multicolumn{1}{c}{(M$\lambda$)} & (mJy) & (mas) & (pc) & (K) & (mJy) &  \\
\multicolumn{1}{c}{(1)} & \multicolumn{1}{c}{(2)} & \multicolumn{1}{c}{(3)} & (4) & (5) & (6) & (7) & (8) & (9) \\\hline
0043$-$001\footnotemark[$*$] & $<$9.9 & 1.5  & \ldots & \ldots & \ldots & \ldots & 103  & \ldots \\
 & $<$3.4 & 4.3  &  &  &  &  &  &  \\
 & $<$4.6 & 5.8  &  &  &  &  &  &  \\
 & $<$2.6 & 17.4  &  &  &  &  &  &  \\
 & $<$5.5 & 21.5  &  &  &  &  &  &  \\
 & $<$7.4 & 22.9  &  &  &  &  &  &  \\
0217$-$005\footnotemark[$\dagger$] & 43  & 1.5  & \ldots & 42--92 & 340--750 & 4.5--5.2 & 212  & $-$0.9 \\
 & $<$3.4 & 4.4  &  &  &  &  &  &  \\
 & $<$4.6 & 5.9  &  &  &  &  &  &  \\
 & $<$2.6 & 18.7  &  &  &  &  &  &  \\
 & $<$5.5 & 22.9  &  &  &  &  &  &  \\
 & $<$7.4 & 24.3  &  &  &  &  &  &  \\
0756+371 & 140  & 3.0  & 139$\pm$14 & 2.0$\pm$0.6 & 16$\pm$5 & 8.9  & 239  & $-$0.3 \\
 & 133  & 10.0  &  &  &  &  &  &  \\
 & 133  & 12.8  &  &  &  &  &  &  \\
0800+402\footnotemark[$*$] & $<$3.4 & 3.3  & \ldots & \ldots & \ldots & \ldots & 190  & \ldots \\
 & $<$2.6 & 11.2  &  &  &  &  &  &  \\
 & $<$5.5 & 14.2  &  &  &  &  &  &  \\
0815+330 & 30  & 2.8  & 31$\pm$4 & 7.5$\pm$0.7 & 61$\pm$6 & 7.1  & 328  & $-$1.3 \\
 & 20  & 9.0  &  &  &  &  &  &  \\
 & 18  & 11.4  &  &  &  &  &  &  \\
0928+444 & 314  & 3.2  & 315$\pm$32 & 1.5$\pm$0.2 & 13$\pm$2 & 9.6  & 156  & $+$0.4 \\
 & 311  & 9.6  &  &  &  &  &  &  \\
 & 305  & 12.7  &  &  &  &  &  &  \\
1005+480\footnotemark[$\ddagger$] & 16  & 3.4  & \ldots & $<$51 & $<$420 & $>$5.0 & 206  & $-$1.3 \\
 & 21  & 10.3  &  &  &  &  &  &  \\
 & 20  & 13.6  &  &  &  &  &  &  \\
1013+491 & 90  & 3.4  & 97$\pm$10 & 8.7$\pm$0.1 & 72$\pm$1 & 7.5  & 267  & $-$0.6 \\
 & 49  & 10.5  &  &  &  &  &  &  \\
 & 28  & 13.9  &  &  &  &  &  &  \\
1018+053 & 554  & 2.6  & 555$\pm$56 & 2.2$\pm$0.1 & 19$\pm$1 & 9.5  & 284  & $+$0.4 \\
 & 530  & 10.5  &  &  &  &  &  &  \\
 & 520  & 12.8  &  &  &  &  &  &  \\
1020+432 & 196  & 3.1  & 234$\pm$92 & 9.9$\pm$2.4 & 83$\pm$20 & 7.8  & 108  & $+$0.3 \\
 & 191  & 9.0  &  &  &  &  &  &  \\
 & 62  & 12.1  &  &  &  &  &  &  \\
1030+085 & 66  & 2.6  & 65$\pm$8 & 3.0$\pm$1.2 & 25$\pm$10 & 8.3  & 108  & $-$0.3 \\
 & 56  & 11.2  &  &  &  &  &  &  \\
 & 59  & 13.5  &  &  &  &  &  &  \\
1042+074 & 131  & 2.6  & 133$\pm$18 & 7.2$\pm$0.5 & 58$\pm$4 & 7.8  & 374  & $-$0.6 \\
 & 62  & 10.5  &  &  &  &  &  &  \\
 & 50  & 12.8  &  &  &  &  &  &  \\
1057+032 & 23  & 2.8  & 23$\pm$3 & 5.8$\pm$0.8 & 42$\pm$6 & 7.2  & 138  & $-$1.0 \\
 & 15  & 11.2  &  &  &  &  &  &  \\
 & 15  & 13.8  &  &  &  &  &  &  \\
1103+023 & 71  & 2.8  & 72$\pm$7 & 4.7$\pm$0.3 & 38$\pm$3 & 7.9  & 163  & $-$0.5 \\
 & 56  & 10.9  &  &  &  &  &  &  \\
 & 52  & 13.4  &  &  &  &  &  &  \\
1119+600\footnotemark[$\S$] & 102  & 3.9  & \ldots & $<21$ & $<170$ & $>$5.5 & 186  & $-$0.3 \\
 & \ldots & \ldots &  &  &  &  &  &  \\
 & \ldots & \ldots &  &  &  &  &  &  \\
\hline
\end{tabular}
\end{center}
\end{table*}

\begin{table*}
\contcaption{({\it continued})}
\begin{center}
\begin{tabular}{lrrcccccc} \hline\hline
\multicolumn{1}{c}{Source} & \multicolumn{1}{c}{$S^\mathrm{cor}_\mathrm{8.4GHz}$} & \multicolumn{1}{c}{$B_{uv}$} & $S^\mathrm{fit}_\mathrm{8.4GHz}$ & $\phi^\mathrm{fit}_\mathrm{FWHM}$ & $\phi^\mathrm{fit}_\mathrm{FWHM}$ & $\log{(T_\mathrm{B})}$ & $I^\mathrm{FIRST}_\mathrm{1.4GHz}$ & $\alpha(I^\mathrm{FIRST}_\mathrm{1.4GHz} - S_\mathrm{8.4GHz}^\mathrm{cor-max})$ \\
\multicolumn{1}{c}{} & \multicolumn{1}{c}{(mJy)} & \multicolumn{1}{c}{(M$\lambda$)} & (mJy) & (mas) & (pc) & (K) & (mJy) &  \\
\multicolumn{1}{c}{(1)} & \multicolumn{1}{c}{(2)} & \multicolumn{1}{c}{(3)} & (4) & (5) & (6) & (7) & (8) & (9) \\\hline
1159+011 & 170  & 3.4  & 169$\pm$17 & $<$0.7 & $<$6 & $>$9.9 & 267  & $-$0.3 \\
 & 168  & 13.3  &  &  &  &  &  &  \\
 & 170  & 16.5  &  &  &  &  &  &  \\
1223+503 & 102  & 3.5  & 105$\pm$15 & $<$2.1 & $<$15 & $>$8.8 & 222  & $-$0.4 \\
 & 113  & 17.5  &  &  &  &  &  &  \\
 & 96  & 21.1  &  &  &  &  &  &  \\
1228$-$030 & 222  & 3.6  & 222$\pm$22 & $<$2.0 & $<$17 & $>$9.2 & 144  & $+$0.2 \\
 & 224  & 13.9  &  &  &  &  &  &  \\
 & 224  & 17.4  &  &  &  &  &  &  \\
1405+405 & 179  & 2.8  & 182$\pm$24 & $<$2.4 & $<$20 & $>$8.9 & 206  & \ \ 0.0 \\
 & 193  & 15.6  &  &  &  &  &  &  \\
 & 170  & 18.4  &  &  &  &  &  &  \\
1432+410 & 45  & 2.9  & 46$\pm$5 & 5.3$\pm$0.1 & 45$\pm$1 & 7.6  & 257  & $-$1.0 \\
 & 26  & 15.6  &  &  &  &  &  &  \\
 & 21  & 18.4  &  &  &  &  &  &  \\
1510+595 & 18  & 3.8  & 19$\pm$3 & $<$2.5 & $<$21 & $>$7.9 & 182  & $-$1.2 \\
 & 20  & 18.1  &  &  &  &  &  &  \\
 & 17  & 21.9  &  &  &  &  &  &  \\
1528+531\footnotemark[$\dagger$] & 12  & 3.6  & \ldots & 7.9--51 & 62--400 & 5.0--6.6 & 172  & $-$1.5 \\
 & $<$2.6 & 17.4  &  &  &  &  &  &  \\
 & $<$5.5 & 21.0  &  &  &  &  &  &  \\\hline
\end{tabular}
\end{center}
{\footnotesize Col.~(1) Truncated name of target source; Col.~(2) Correlated flux density; Col.~(3) Projected baseline length; Col.~(4) Fitted flux density of a Gaussian profile (see Sections~\ref{section:observationanddatareduction} and \ref{section:result}); Col.~(5) Fitted FWHM size of a Gaussian profile; Col.~(6) Fitted FWHM in pc; Col.~(7) Brightness temperature calculated using Eq.~(\ref{equation:brightnesstemperature}); Col.~(8) FIRST 1.4~GHz peak intensity, the same as in Table~\ref{table1}; Col.~(9) Spectral index between 1.4 and 8.4~GHz, calculated from the FIRST peak intensity and the maximum correlated flux density.\\
\footnotemark[$*$] Undetected at all baselines, indicating a $7\sigma$ upper limit of the shortest baseline.\\
\footnotemark[$\dagger$] Detected only at the shortest baseline.\\
\footnotemark[$\ddagger$] Complex visibility profile; cannot be fitted with a Gaussian.\\
\footnotemark[$\S$] Only a single baseline observation.}
\end{table*}

\section{Results}\label{section:result}
With OCTAVE baselines, we detected 20 of the 22~FIRST/SDSS BAL quasars.  Our OCTAVE observations provided baseline detection limits to brightness temperatures, $T_\mathrm{B}$, of $10^5$--$10^6$~K for a $\sim9$-minute scan period.  Brightness temperatures were calculated from 
\begin{equation}
T_\mathrm{B} = 1.8 \times 10^9 (1+z) \frac{S_\nu}{(\nu^2 \phi^2)}
\label{equation:brightnesstemperature}
\end{equation}
in Kelvin, where $z$ is redshift, $S_\nu$ is the flux density in mJy at frequency $\nu$ in GHz, $\phi$ in mas is the fitted FWHM of the source size (cf.~\cite{Ulvestad_etal.2005}).   We have derived the resolved sizes for 11~sources at mas scales ranging 1.5--10~mas, implying $T_\mathrm{B}=10^{7.1}$--10$^{9.6}$~K.  Five sources were unresolved~($\lesssim 2$~mas) with the OCTAVE baselines, implying $T_\mathrm{B} \gtrsim 10^{8}$~K.  The sensitivity in the determination of the brightness temperature for each source depends on the projected baseline lengths, signal-to-noise ratio, uncertainty in amplitude calibration, and any discrepancy between the two-dimensional intrinsic source profile and the one-dimensional fitting model (see Section~\ref{section:observationanddatareduction}).  For FIRST~J100515.9$+$480533, the visibility profile cannot be represented by a simple Gaussian, which may suggest the existence of multiple components with comparable fluxes at mas scale.  A complicated jet structure was inferred in this BAL quasar.

We also show two-point spectral indices between the FIRST 1.4-GHz peak intensities and the maximum correlated flux densities of the OCTAVE at 8.4~GHz (Table~\ref{table3}).  We used the peak intensities rather than total flux densities at 1.4~GHz to extract emitting components as close to the nuclei as possible.  The FIRST data were obtained with a $\sim$\timeform{5''}-resolution, which is much larger than that of the OCTAVE; thus, the resolution effect could occur in deriving the spectral indices.  Note that the derived spectral indices provide only lower limits, if any of the radio emitting components in a source lies outside a region within $\sim$10~mas.  For this reason, we can be confident of the detections of inverted spectra ($\alpha > 0$).  On the other hand, we cannot determine whether steep spectra ($\alpha < 0$) were the results from an intrinsic nature or a resolution effect.  In the 22 targets, we found inverted spectra in four sources.  Note that the derivations of spectral indices could also be affected by flux variability, as $\sim10$~years elapsed between the FIRST and OCTAVE observations.  We discuss this possibility later.

\section{Discussion}\label{section:discussion}
\subsection{Coexistence of nonthermal jet and BAL outflow}\label{section:coexistence}
We detected 20 radio sources using OCTAVE baselines, which assured brightness temperatures of greater than $10^5$--$10^6$~K.  Although the brightness temperature of a radio supernova at a very early stage could exceed $T_\mathrm{B}=10^6$~K (e.g., \cite{Bietenholz_etal.2001}), its expected flux density would be far from sufficient for fringe detection for such distant quasars in our sample (cf.~\cite{vanDyk_etal.1993}).  The brightness temperatures of stellar components of luminous starbursts are $\lesssim10^5$~K (cf. \cite{Lonsdale_etal.1993}), which are less than those of the detected radio counterparts of BAL quasars.  Also, in terms of radio luminosity, even the most radio-luminous starbursts show up to only $\sim10^{24}$~W~Hz$^{-1}$ (e.g., \cite{Smith_etal.1998}), in contrast to $\sim10^{27}$--$10^{28}$~W~Hz$^{-1}$ for our BAL quasar sample.  Therefore, the OCTAVE-detected radio emissions cannot be accounted for by any stellar origin.  We conclude that the OCTAVE-detected radio emissions of the BAL quasars originate in nonthermal jets from AGN activity, as in the cases of other AGN radio sources.  VLBI observations that previously conducted \citep{Jiang&Wang2003,Kunert-Bajraszewska&Marecki2007,Liu_etal.2008} also support the existence of nonthermal jets in BAL quasars.  

The fairly high VLBI-detection rate (20/22) is evidence that BAL outflows, which are inferred from broad troughs in UV spectra, can coexist with nonthermal jets in radio-loud BAL quasars.  This indicates that the accretion disks of BAL quasars can generate radiation-pressure driven strong outflows and magnetic-driven strong jets simultaneously.  It is important to investigate the relationship between the properties of BAL features and radio emissions (cf.~\cite{Ghosh&Punsly2007}) to understand accretion phenomena in quasars.

\subsection{Jet properties of observed BAL quasars --- inverted-spectrum sources}\label{section:jetproperties:invertedspectrum}
We found four inverted-spectrum ($\alpha > 0$) sources (Column~(9) in Table~\ref{table3}).  Since optically-thin nonthermal synchrotron emission should show steep spectrum of $\alpha \leqq -0.5$, the observed inverted spectra should result from a lower-frequency absorption mechanism if a non-simultaneous spectral index is not affected by flux variability.  Absorbed components should dominate the total spectrum for an inverted spectrum to be observed even using the FIRST beam width.  Hence, an inverted spectrum suggests three possibilities: (1)~Doppler beaming effect on jets, (2)~a young~(compact) radio source, or (3)~artificially made due to flux variability, as follows.  

Doppler boosting can apparently enhance only optically-thick nuclear components beyond extended jets that would have been decelerated and optically-thin: the nuclear components tend to be synchrotron self-absorbed because of high-brightness temperatures, and show inverted spectrum at lower frequencies.  An adequate Doppler beaming effect requires jets that are nearly aligned with our line-of-sight; this implies a face-on viewed accretion disk, which is inconsistent with the widely accepted scheme of equatorial BAL outflows.  The presence of Doppler beaming effect has already been inferred from rapid flux variation between the NRAO VLA Sky Survey (NVSS; \cite{Condon_etal.1998}) and the FIRST in several BAL quasars \citep{Zhou_etal.2006,Ghosh&Punsly2007}.  It is important to confirm the Doppler beaming in also by VLBI imaging in a future.  

Young radio sources also can make their spectra nuclear-dominated, because extended jets would not yet been developed.  Radio sources of $\sim100$--1000~pc or less could show peaked spectra, resulting inverted spectra possibly at $\sim1$~GHz; the spectral-peak frequency is roughly determined by the linear size of radio structure (e.g., \cite{Snellen_etal.2000}).  The explanations of the relation between linear size and spectral-peak frequency have been suggested using synchrotron self-absorption~\citep{Odea&Baum1997} and free--free absorption~\citep{Bicknell_etal.1997}.  \citet{Montenegro-Montes_etal.2008} presented many radio sources of BAL quasars showing such peaked spectra.  The ages could be determined from the linear size and expanding speed of radio structures; it is important to measure the expanding speed by VLBI-imaging monitor in a future.  For example, in a typical VLBI scale, an apparent expanding rate of 0.1~mas yr$^{-1}$ and a linear size of two-sided structure of 100~pc leads to an estimated age of 500~yr.  This situation can be adapted to the evolutionary scenario of BAL quasars in a relatively short-lived phase (e.g., \cite{Briggs_etal.1984,Gregg_etal.2000}, and see also \cite{Gregg_etal.2006}).

Inverted spectra artificially-made due to flux variability between the observations of the OCTAVE at 8.4~GHz and the VLA at 1.4~GHz ($\sim10$~yr) cannot be ruled out.  A fraction of BAL quasars are significantly variable in the radio band \citep{Zhou_etal.2006,Ghosh&Punsly2007}.  A flux variability of $>$10\% in 10~yr at 8.4~GHz for a 100-mJy source at $z=2$ implies $T_\mathrm{B} > 10^{11.3}$~K, which would be above the inverse-Compton limit \citep{Readhead1994} and require Doppler beaming effect.  A flux variability of 10\% corresponds to a change of spectral index of only $\sim0.05$.  That means that even if the observed spectra had been artificially made, we have the same conclusion that the four inverted-spectrum sources are possibly Doppler-boosted.  {\it As a result, we conclude that the inverted-spectrum sources are interpreted as Doppler-beamed, pole-on-viewed relativistic jet sources or young radio sources such as CSSs and GPSs: single edge-on geometry cannot descibe all BAL quasars.}  

We also check flux variation between NVSS and FIRST at 1.4~GHz for our BAL quasar sample.  Using the same manner as \citet{Ghosh&Punsly2007}, we found two sources, FIRST J075628.2$+$371455 and FIRST J122836.9$-$030438, with significant flux variation of $4.8\sigma$ and $3.2\sigma$; brightness temperatures were derived to be $10^{15.3}$~K and $10^{13.2}$~K, respectively.  Both of the radio sources were very compact (2~mas or less) in OCTAVE observations, and the latter showed an inverted spectral index ($\alpha=+2.0$).  These situations are consistent with the presence of highly Doppler boosting on pole-on jets.  It is important to confirm them also by VLBI imaging in a future.

\subsection{Jet properties of observed BAL quasars --- steep-spectrum sources}\label{section:jetproperties:steepspectrum}
The majority of the detected BAL quasars showed steep ($\alpha<0$) radio spectra.  It was unclear whether the weaker radio flux densities were observed at 8.4~GHz because extended structures were resolved out, or were due to intrinsically steep spectra on compact components.  Hence, we do not discuss further in detail for the OCTAVE results.  We have already stressed the importance of VLBI imaging observations for inverted-spectrum sources, and steep-spectrum sources are also needed to be VLBI-observed because they are the majority of BAL quasars and are also compact in most cases \citep{Becker_etal.2000}.  The correlated flux densities in the OCTAVE baselines were not much smaller than the $10^{-1}$-times the VLA peak intensities in most sources, despite the fact that the beam area of the OCTAVE was $\sim10^{-5}$-times that of VLA.  This indicates that radio-emitting origins considerably concentrated in parsec-scale components, which should be investigated using VLBI.  As the evolutionary scenario, BAL quasars might associate young radio sources \citep{Briggs_etal.1984,Gregg_etal.2000,Gregg_etal.2006}, which should be optically-thin small radio lobes seen in compact steep spectrum~(CSS; \cite{Odea1998} for a review) objects.  Many CSS sources have been revealed as young ($<10^5$~yr) radio galaxies by VLBI-imaging studies (e.g., \cite{Murgia2003,Nagai_etal.2006}).  VLBI-imaging studies may provide evidence supporting the evolutionary scenario for BAL quasars, if the steep-spectrum radio sources of BAL quasars are CSS objects \citep{Kunert-Bajraszewska&Marecki2007}.

\subsection{Implications of OCTAVE observations}\label{section:implications}
Our OCTAVE observations have many implications for the study of BAL quasars.  VLBI images of only four BAL quasars have been published before the OCTAVE observations \citep{Jiang&Wang2003,Kunert-Bajraszewska&Marecki2007}.  Our OCTAVE observations have dramatically increased the number of VLBI-detected BAL quasars, and have established that this AGN subclass includes a non-trivial number of radio-loud objects that can be directly imaged at mas resolutions (corresponding to parsec scales) as well as objects in the other AGN classes.  The orientation angle and the age as radio sources of BAL quasars should be determined from jet properties in subsequent detailed VLBI imaging studies.  The OCTAVE strongly recommends that multi-epoch and multi-frequency (and polarimetric) VLBI observations should be performed for these sources.  Multi-epoch imaging can measure the proper motion of jets or lobes to estimate the inclination or kinematic age.  Multi-frequency imaging can measure the spectral-index profiles along approaching and receding jets or lobes, which can discriminate between Doppler-boosted self-absorbed jets and free--free absorbed jets obscured by thermal plasma.  BAL outflows could be free--free absorber along nonthermal jets; the projected one-dimensional profile of thermal BAL outflows could be studied in mas resolutions using VLBIs at multi-frequency at by measuring opacity gradient along the jets.  The free--free absorber also could occur Faraday rotation to the vector of polarization axis toward jets, which is an another powerful tool to investigate the spatial profile of thermal BAL outflows.  In multi-frequency VLBI observations, such processes should offer exclusive probes to the parsec-scale profile of BAL outflows.  On the basis of the results of the OCTAVE observations, we are observing some of the OCTAVE sample by multi-frequency VLBI imaging.  It will be important to compare the results of inclination measurements based on the VLBIs and optical spectropolarimetry, and to investigate the relation among UV/optical spectra, pc-scale geometry, and ages.

\section{Summary}\label{section:summary}
Two of the most important questions in understanding BAL quasars are: (1)~Are they viewed at nearly edge-on? (2) Are they in a short-lived evolutionary phase?  Our OCTAVE observations have increased the number of VLBI-detected BAL quasars, which offers more chance to determine their orientations and ages by subsequent VLBI-imaging studies to conclude the two pictures.  We detected 20 out of the radio-brightest 22~sources selected from the counterparts of SDSS BAL quasars.  We concluded that nonthermal pc-scale jets and thermal BAL outflows can coexist in these radio-loud BAL quasars simultaneously.  BAL quasars have become targets of VLBIs to be revealed in pc scale by direct imaging, as in the cases of other AGN classes.  We also found four inverted-spectrum sources, which are interpreted as Doppler-beamed, pole-on-viewed relativistic jet sources or young radio sources: single edge-on geometry cannot describe all BAL quasars.

\bigskip

% Acknowledgments
We thank M.~S.~Brotherton for very useful comments that improved the presentation of the paper.  We also thank K.~Asada for very useful comments.  This work was partially supported by a Grant-in-Aid for Young Scientists~(B; 18740107, A.~D.), a Grant-in-Aid for Scientific Research~(C; 21540250, A.~D.) and a Grant-in-Aid for Scientific Research~(C; 20540233, K.~W.) from the Japanese Ministry of Education, Culture, Sports, Science, and Technology~(MEXT).  We are grateful to all the staffs and students involved in the development and operation of the OCTAVE.  The OCTAVE project has been developed as a subproject under the VLBI Exploration of Radio Astrometry~(VERA) project in the National Astronomical Observatory of Japan~(NAOJ), a branch of the National Institutes of Natural Sciences~(NINS).  The optical-fiber networks for OCTAVE have been provided by the GALAXY project\footnote{The GALAXY lines connecting Usuda 64~m and Nobeyama 45~m to correlators had been closed since April 2008.} supported by NTT Corporation \citep{Uose_etal.2002}, the Japan Gigabit Network-2 (JGN2) project operated by the National Institute of Information and Communications Technology~(NICT), and the Science Information NETwork~3~(SINET3) operated by the National Institute of Informatics~(NII).  The OCTAVE array consists of six contributed antennas: the Usuda 64~m of the Japan Aerospace Exploration Agency~(JAXA), the Kashima 34~m of the NICT, the Nobeyama 45~m of the Nobeyama Radio Observatory~(NRO) in the NAOJ, the Tsukuba 32~m of the Geographical Survey Institute~(GSI), the Yamaguchi 32~m (operated by Yamaguchi University) of the NAOJ, and the Gifu 11~m of Gifu University.  Time allocation and array operation for OCTAVE are being carried out under the framework of the Japanese VLBI Network~(JVN) project, which is led by the NAOJ, Hokkaido University, Gifu University, Yamaguchi University, and Kagoshima University, in cooperation with the Institute of Space and Astronautical Science~(ISAS)/JAXA, GSI and NICT.  We used the US National Aeronautics and Space Administration's (NASA's) Astrophysics Data System~(ADS) abstract service, the NASA/IPAC Extragalactic Database (NED), which is operated by the Jet Propulsion Laboratory~(JPL).  In addition, we used the Astronomical Image Processing System~(AIPS) software developed at the US  National Radio Astronomy Observatory~(NRAO), a facility of the National Science Foundation operated under cooperative agreement by Associated Universities, Inc.

% \clearpage
% \renewcommand{\arraystretch}{0.5}

\end{document}